\documentclass[10pt]{scrartcl}

\usepackage[utf8]{inputenc}
\usepackage[T1]{fontenc}

\usepackage[natbibapa]{apacite}
\usepackage{amsmath}
\usepackage{amsfonts}
\usepackage{amssymb}
\usepackage{lmodern}
\usepackage{color}
\usepackage{graphicx}
\usepackage[lmargin=2cm,rmargin=2.5cm]{geometry}
\usepackage{lscape}
\usepackage{multicol}
\usepackage{subcaption}
\usepackage{empheq}
\usepackage{gensymb}
\usepackage{hyperref}
\usepackage{float}
\usepackage{csquotes} 
\usepackage{tabularx}
\usepackage[table]{xcolor}
\usepackage{booktabs}
\RequirePackage{silence}
\usepackage{array}
\newcolumntype{x}[1]{>{\centering\let\newline\\\arraybackslash\hspace{0pt}}p{#1}}

\usepackage{setspace}

\begin{document}
\onehalfspacing 


\title{Distinctive pupil and microsaccade-rate signatures in self-recognition}

\author{Lisa~Schwetlick$^{1,3,6, *}$ \and Hendrik~Graupner$^{2,4}$ \\ \and Olaf~Dimigen$^{5}$ \and  Ralf~Engbert$^{1,3}$}



\maketitle

\noindent 1 Department of Psychology, University of Potsdam, Germany \\ 2 Bundesdruckerei GmbH, Berlin, Germany \\ 3 DFG Collaborative Research Center 1294, University of Potsdam, Germany \\ 4 Hasso Plattner Institute and University of Potsdam, Germany \\ 5 Faculty of Behavioural and Social Sciences, University of Groningen, Netherlands \\ 6 Laboratory of Psychophysics, EPFL, Lausanne, Switzerland \\ * lisa.schwetlick@epfl.ch

\vspace{1cm}

\begin{abstract}
Pupil dynamics and fixational eye movements are \textcolor{black}{primarily} involuntary processes that actively support visual perception during fixations. Both measures are known to be sensitive to ongoing cognitive and affective processing. In a visual fixation experiment ($N$=116) we demonstrate that self-recognition, familiar faces, and unfamiliar faces elicit specific responses in pupil dynamics and microsaccade rate.
First, the pupil response comprises an immediate pupil constriction followed by a dilation in response to stimulus onsets.
We observe attenuated constriction and greater dilation when faces are recognized compared to unknown faces.  This effect is strongest for one's own face.
Second, microsaccade rates, which generally show inhibitory responses to incoming stimuli, generate stronger inhibition for familiar faces compared to unknown faces. Again, the strongest inhibition is observed in response to one's own face.
Our results imply that eye-related physiological measures expose hidden knowledge in face memory and could contribute to biometric authentication and identity validation procedures.
\end{abstract}

\newpage

\section{Introduction}
Eye movements are an essential component of visual perception. Easy to observe are rapid eye movements (saccades) that generate sequences of gaze shifts and periods of relative rest (fixations) that give rise to the eye's trajectory during scene exploration \citep{Henderson2003} or everyday activities. In the laboratory, we can also investigate microscopic eye movements that occur during fixations: microsaccades and changes in pupil size. Here we investigate how both systems are influenced by face recognition.

Microsaccades are generated by the same machinery of extraocular muscles that drive gaze shifts. These small-amplitude fixational eye movements share their kinematic relations with larger saccades \citep{Bahill1975}. Microsaccades are distinguished from the larger saccades primarily by their size  \citep[but not exclusively, see][]{Mergenthaler2010}, typically <1$^\circ$ of visual angle. During fixations, they serve vital functions for visual processing, e.g., counteracting perceptual fading \citep{MartinezConde2004}\textcolor{black}{, and support visual exploration at the foveal scale \citep{Shelchkova2019}. Microsaccades were initially defined as involuntary motor behavior \citep{Ratliff1950} during steady fixation, but during high-acuity tasks microsaccades might be better described as primarily involuntary with voluntary components \citep{Sinn2016,Willeke2019}.}

The second type of investigated eye movement, pupil dilation, driven by intraocular muscles and primarily adapts pupil size to the surrounding illumination (pupillary light reflex or PLR) \citep{Mathot2018}. Both microsaccades and pupil dilation are modulated by ongoing cognitive \citep{Engbert2006a} and emotional processing \citep{Hess1960,Winterson1976}. In the current study we set out to investigate how the recognition of unfamiliar and familiar faces--including one's own face--affects microsaccades and pupil size.

Face recognition is tightly coupled to visual, cognitive and affective processing. Evidence from the literature suggests that faces are processed in the visual system by unique, specialized cognitive and neural mechanisms \citep{McKone2005, Yin1969} in the fusiform gyrus of the inferior temporal (IT) cortex \citep{Kanwisher1997} and in prefrontal cortex \citep{Rolls2005}. Response latency for face-sensitive cells in IT cortex is reported to be between 50~ms \citep{Sugase1999}  and  160~ms \citep{Perrett1982}.
\textcolor{black}{Moreover, recent findings highlight the role of the superior colliculus (SC) in detecting objects, including faces, at very early stages of visual processing. SC neurons respond robustly to facial stimuli and face-like patterns as little as 50 ms after stimulus onset \citep{Nguyen2012, Nguyen2014, Nguyen2016}. Specifically, \citet{Bogadhi2023} showed that SC neurons exhibit immediate visual bursts that differentiate objects from non-objects, pointing toward its role in rapid, coarse visual categorization.}

The recognition of a specific person, or even oneself, engages the same visual processing circuitry as a new, unfamiliar face, but additionally causes memory retrieval, context-dependent surprisal, and emotional responses \citep{Johnston2009}. In order to isolate visual processing components of recognition, previous studies typically used faces that entrained during the experiment (unfamiliar face recognition) or famous faces (familiar face recognition). We expect that semantic and emotional responses will be significantly stronger for faces of real-life acquaintances and friends, as well as oneself (self-recognition).

Pupil size is a well-established physiological measure of ongoing cognition that varies between 1.5~mm to 9~mm in diameter \citep{Sirois2014}. Modern pupillometry distinguishes three types of pupil response: The pupilary light reflex (PLR), pupil near response (PNR), and psychosensory pupil responses \citep{Mathot2018}. Of these, the PLR is the strongest component, causing pupil constriction within the first 200~ms after stimulus onset; the minimum pupil size is reached between 200~ms \citep{Mathot2018} and 1600~ms \citep{Bradley2008}.
In addition to external factors, pupil size is also determined by cognition and arousal, modulated by variables such as interest or processing load. Physiologically, pupil size is connected to brain regions that control sleep-wake rhythms and nervous-system activation \citep{Sirois2014}. Activity in the Locus Coeruleus, which is involved in memory retrieval and selective attention, is highly correlated with pupil size in monkeys \citep{Laeng2012}.
Early cognitive effects on pupil size (within 1~s) are related to novelty and saliency \citep{Mathot2018, Sirois2014, Hess1960} as well as surprisal, uncertainty, and prediction errors \citep{Larsen2018}. The effects of (positive and negative) arousal and mental effort are associated with longer delays \citep[see][for a review]{Mathot2018}, while emotional responses peak last (>2~s after target onset)  \citep{Kinner2017}. 

For microsaccades during cognitive tasks, the baseline rate is usually around 1 per second (1~Hz). Changes in visual input, however, can strongly modulate the rate \citep{Engbert2003,Engbert2012,Martin2020}.
In particular, any change in the display causes a temporary decrease in the microsaccade rate (MSR). The microsaccadic inhibition is followed by a rise in MSR, often exceeding the baseline rate. 500-1000~ms after stimulus onset the MSR returns to its resting state.
\textcolor{black}{While early research suggested that display-change-related microsaccade inhibition is generated at the level of the superior colliculus (SC) \citep{Laubrock2013, Rolfs2009}, more recent evidence shows that saccadic inhibition is unaffected by SC inactivation \citep{Hafed2013}. As a generating mechanism, \citet{Hafed2013} proposed that microsaccadic inhibition results from phase resetting of ongoing oscillatory motor programs. Furthermore, \citet{Buonocore2021, Buonocore2023} highlighted the role of brainstem mechanisms, such as omnipause neurons, in initiating the transient cessation of saccades in response to visual transients, effectively serving as a ``brake'' that resets motor plans. The SC receives input from various regions potentially conveying top-down information  \citep{Sparks2002, Schall1995}, responds specifically to real-life objects like faces with early visual bursts \citep{Nguyen2012, Nguyen2014, Nguyen2016, Bogadhi2023} and may influence microsaccades indirectly. In summary, both previous behavioral data and our current understanding of the underlying neurophysiology suggest that pupil size and MSR could vary substantially across conditions during face recognition.}

Recent studies have explored pupil dilation and MSR in the context of face recognition, using a paradigm measuring reactions to rapid presentation of faces. \citet{Chen2023} showed streams of unknown faces, target faces, and a familiar face (the participants mother) and found that pupil dilation is modulated by face recognition, but found no effects on MSR.
\citet{Rosenzweig2019} specifically probed for microsaccadic inhibition using a similar rapid presentation paradigm, where they compared (1) a pre-learned target face (the \enquote{terrorist}) within a set of 8 unknown faces \citep{Rosenzweig2019} or (2) a famous faces among unknown faces \citep{Rosenzweig2020}. They found evidence of stronger microsaccadic inhibition related to face familiarity.

The goal of the present work was to establish in a large-scale eye-tracking experiment that pupil dilation and microsaccade dynamics reveal face recognition, and more specifically also self-recognition (as an extreme example of a familiar face). We manipulated face familiarity by using pictures of the participants classmates, yielding a well-controlled design with a high statistical power. We were also interested in the influence of image repetition, individual differences, the time-course of both measures, which are highly relevant for potential applications.
Compared to unfamiliar faces and pre-learned faces, real-life familiar faces should be intrinsically relevant and meaningful to the observer, even in the absence of a specific task. We therefore hypothesized that recognition should produce reliable oculomotor signature, akin to an oddball response. In an oddball paradigm, participants see a sequence of frequent (expected) stimuli, interspersed with infrequent (and therefore unexpected) but task-relevant targets (the oddball stimulus) \citep{Sutton1965}, while performing a task such as silently counting targets. This paradigm \citep[see][for an extensive review]{Polich2007} has revealed that a particular event-related potential (P300, peaking around ~300 ms after presentation) is strongest when the participant is engaged in target detection \citep{Squires1975, Sutton1965,Picton1992} and is modulated by the meaningfulness of stimuli \citep{Johnson1992}, task difficulty \citep{Picton1992}, as well as the internal arousal level and availability of attentional resources \citep{Kok2001}.

The oddball task has been found to affect pupil dilation \citep{Kamp2015, Strauch2020, Murphy2014}, likely via the Locus Coeruleus \citep{Murphy2014}, as well as MSR \citep{Valsecchi2006,Valsecchi2009}. 
We expect face recognition to comprise various reflexive, cognitive, and emotional components and to be similar to the surprisal and increased attention modulated via target-frequency in the oddball task.
 The recognition of familiar faces should be related to a reduced MSR and increased pupil dilation, compared to unfamiliar faces. Moreover, we expect recognition of one's own face to evoke the strongest response, compared to familiar and unfamiliar faces.

From the literature reviewed above, we conclude that both pupil size and MSR are physiological measures that are functionally-driven but modulated by ongoing cognition. The aim of the present study is to investigate whether pupil dynamics and MSRs are specifically affected by face recognition processes. Reliable differences in pupil dynamics and microsaccade statistics between self-recognition and recognition of other faces are relevant to applications in the context viewer identification or exposing viewer knowledge.


\section{Methods}
\subsection{Participants}
We recruited participants from two graduating high school classes in Potsdam, Germany. The advantage of this setup is that it allows a well-controlled design where each participant personally knows a subset of the other participants. 
Thus, each participant was shown faces that were either their classmates' faces, who they knew personally (\emph{Peers}), or students from the other high school (\emph{Strangers}), or their own (\emph{Self}). Of the initial cohort of 127 students who's pictures were taken, 118 students came to the eye-tracking session. Of these, two individuals were not able to be calibrated in the eye tracker. 116 complete data sets remain. The participants were between the ages of 16 and 18 years old; 55 were male, 63 female and one identified as non-binary.

The eye tracking data collection was followed by a further questionnaire, where each participant saw the same faces they had seen during the experiment and was asked to confirm whether they did indeed know the person in each picture. Cases where a participant's response to the question did not match with the expectation, i.e.,  when they did know someone from the other class or did not know someone from their own class, were removed from the final data set. The final data set is therefore balanced in the sense that each image, in principle, appears in each condition.

\subsection{Photographs}
In order to collect the required photographs of the participants, a professional photographer visited the schools. Each participating student received an anonymous code. The photographs were taken under consistent illumination and with the heads centered, in a quasi-biometric setup, in order to maximize consistency between the pictures. The photographs were coded with the anonymous code, so that the mapping between participant and photograph was possible without the necessity of saving any identifying information alongside the image. The high-resolution photographs were further cropped and scaled, to center each face inside a square image. For data-privacy reasons the photographs of the faces were deleted upon completion of the study. In the data set only the anonymous codes remain. 

We computed the relative image luminance for each image and found it to be normally distributed according to a Shapiro Wilk \citep{Shapiro1965} test of normality (\textit{p=0.24}). As each image is shown in each condition (i.e.,  every image is seen as Peers, Strangers, and Self), differences in image statistics would effect all groups equally and not influence the main effects. 

\subsection{Eye Tracking Data Collection}
For recording eye trajectories we used an Eyelink 1000 eye tracker  (SR Research, Osgoode, ON, Canada), which recorded both eyes at 500~Hz at an illumination level of 75\%. The screen,a ViewPIXX monitor (VPixx Technologies, Saint-Bruno, Canada) with a resolution of 1920$\times$1080~pixels, was placed at 70~cm distance from the participant, with the head stabilized in a chin rest. The eyes were centered at $3/4$ of the height of the screen. We used a 5-point calibration grid and subjects were re-calibrated every 14 trials. As the experiment required exclusively fixation in the center (and no exploration of the outlying regions of the screen), a 5-point calibration was sufficient to ensure data quality.

\begin{figure}[t]
    \centering
    \includegraphics[width=0.8\textwidth]{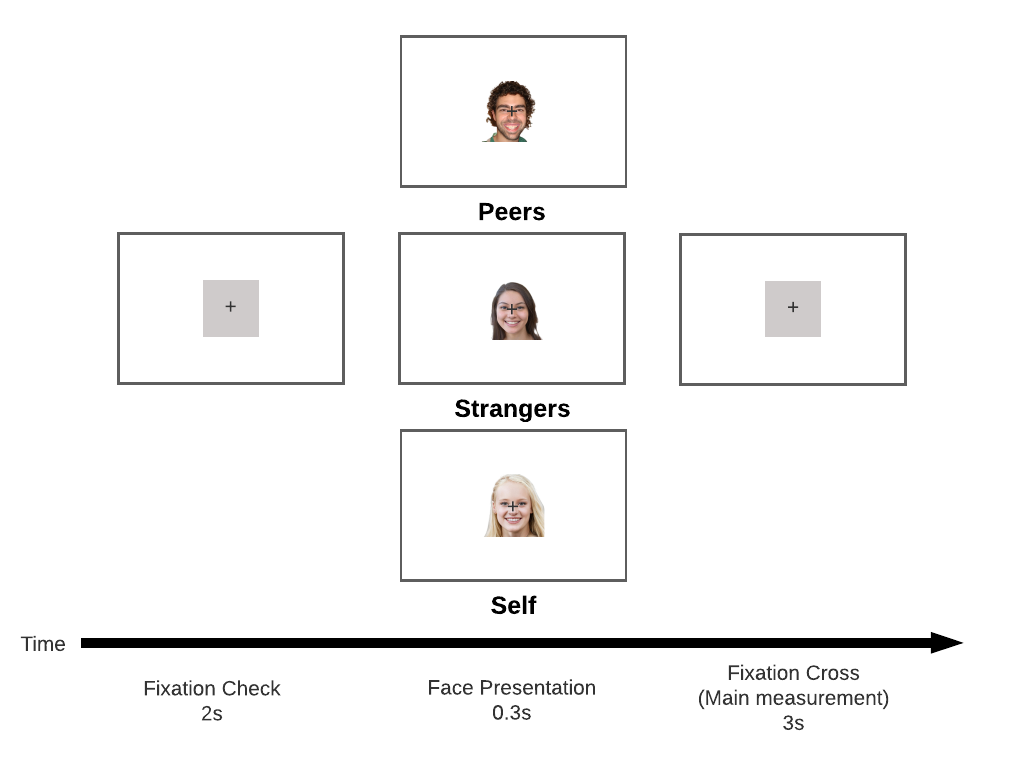}
    \caption{Time course of a trial. The 2~s fixation check ensures participants fixate the center of the screen. The brief face presentation (0.3~s) is then followed by a grey screen, during which the main measurement is taken.}
    \label{fig:trial}
\end{figure}

Face images were shown at a resolution of 500$\times$500~pixels centered on the screen, meaning that each image subtended 11.4$^\circ$ of visual angle. The experimental session proceeded as follows. Participants were informed that they would be seeing pictures of faces and were asked to fixate the central fixation marker for the duration of the trial. 
No task was given apart from fixating the fixation marker. 
The face photograph appeared under the fixation marker for a duration of 300~ms. They were asked not to blink and not to move their eyes. Figure \ref{fig:trial} shows the time course of a single trial. 

The experiment was preceded by three training trials, in order to let participants get used to the procedure. The three faces for the training block were taken from the same image data set but were not used again for trials of the same subject. After the training block, the experimental trials followed. Each trial takes roughly 5~s, split into three phases: (a) A fixation check of around 2~s, i.e.,  if the eyes are not centered on the marker a re-calibration is initiated, (b) an interval of 300~ms of presentation of the face photograph, and (c) a fixation cross for 3~s. After each trial participants were encouraged to take a break and to blink. Using a key-press participants indicated when they were ready for the next trial.

Phases a) and c) showed a grey background which covered the same area as the following image. The shade of grey in the background was designed to be adjusted to the luminance of the stimulus. The luminance was proportional to, but not equal to the stimulus luminance, ranging from a relative luminance of 3.5 to 6~\% while the images varied in relative luminance from 2.5 to 6.5~\%. As the distribution of images was well controlled by the experimental design, i.e.,  each image appeared in each condition, any differences between images can be accounted for by random effect of image in a linear mixed model (LMM) Analysis.  

Each subject saw the following 100 trials in random order:
{\renewcommand{\theenumi}{\alph{enumi}}
\begin{enumerate}
\itemsep0ex
\item 10 repetitions of a photograph of their own face (10 trials)
\item 5 repetitions of photographs of 3 selected peers (15 trials)
\item 5 repetitions of photographs of 3 selected strangers (15 trials)
\item 1 photograph each of 30 strangers (30 trials)
\item 1 photograph each of 30 peers (30 trials)
\end{enumerate}
}
Categories b and c were introduced in order to control for repetition effects that occur during the repeated viewing of one's own face. In order to exclude a pure oddball effect for the Peers-Strangers every participant saw is an equal number of familiar and unfamiliar faces.

\subsection{Pupillometry Analyses}
In order to prepare the data for analysis we closely followed the recommended pre-processing pipeline suggested by \citet{Mathot2022}. Pupil size data was down-sampled to 100~Hz, as a higher resolution is not informative for pupil responses. Missing data were linearly interpolated (and excluded from the analyses). We converted the pupil diameter (which is given by the Eye tracker in arbitrary units) to mm as our base unit and then computed the pupil response of each trial relative to the baseline. The baseline value for each trial was the average pupil size during the 50~ms surrounding target onset (as proposed by \citet{Mathot2022}). We ensured data quality by evaluating the pupil size during the pre-stimulus phase. We found no indication that any trials or participants had to be excluded on the basis of the baseline values. Trials that included a blink after stimulus onset were excluded from the analyses. 

First, as a qualitative analysis, we plotted the pupil response over time,  normalized to the stimulus onset. Second, to statistically support our findings we use linear mixed models (LMM) with pupil size as the dependent variable. While analyses that more efficiently take advantage of the the time series data, such as cluster-based permutation tests, are available and suggested for pupillometry data \citep{Mathot2022}, they were not feasible for this use case, as our design had large differences in the number of samples per group (the Self, First Presentation condition exists only a single time for each participant, but many times for Strangers). We therefore chose to analyze both measures using LMMs in 5 time windows: Baseline (-50 -- +50~ms), Constriction (200 -- 600~ms), Dilation (600 -- 1200~ms), Late Dilation (1200 -- 2000~ms), and Stability (2000 -- 3000~ms). The baseline window is included as a sanity check, to ensure no effects are found before target onset.

Within each window we computed LMMs for the dependent variable "Pupil Size" using the lme4 package \citep{Bates2015a} in R \citep{RCore2021}.
We define the following (custom) contrasts \citep{Schad2020} in our LMM, to test our hypotheses: 
\begin{enumerate}
    \item We compare the Self condition to the Strangers and Peers condition, resulting in a Self--Others comparison.
    \item  We compare the Strangers and Peers conditions (Strangers--Peers).
    \item We compare the repetitions of individual images up to the 5th repetition, using a sliding difference contrast, meaning that we compare presentation 1 to presentation 2, presentation 2 to presentation 3 and so forth.
    \item We compare the interaction of Repetition and Face, which indicates whether the difference in the face conditions is different based on Repetition.
\end{enumerate}

Adding Time within the window and the ordinal trial number as covariates yields the following fixed effects structure for the model formula:

\begin{equation}
Pupil\_Size \sim Face * Repetition + Time + Trial
\end{equation}

The selection process for the random effect structure, is described in Appendix \ref{appx:sec:lmm}. Following \citet{Baayen2008}, we interpret all |t| > 2 as significant fixed effects.

\subsection{Microsaccade Rate Analyses}
Microsaccades were detected with millisecond accuracy from raw data by applying a standard microsaccade detection algorithm using a velocity threshold \citep{Engbert2003,Engbert2006a}. MSR can be estimated with the help of a response function that applies filter kernels to the series of onset times \citep{Engbert2021,Engbert2006a}. From the series of microsaccade onset times $\{t_1, t_2, t_3, ...\}$, the microsaccade rate $r(t)$ at time $t$ is estimated via \begin{equation}
r_{\mathrm approx}(t) = \int_{-\infty}^{+\infty} {\mathrm d}\tau w(\tau)\rho(t-\tau) \;,
\end{equation}
where the microsaccadic response function $\rho(t)$ \citep{Engbert2021} is defined as 
\begin{equation}
\rho(t)=\sum_{i=1}^N \delta(t-t_i) 
\end{equation}
with Dirac's $\delta$-function $\delta(t)$. We applied a filter kernel known as a causal window, i.e.,  
\begin{equation}
w(\tau)=\left[ \alpha^2\tau\exp(-\alpha\tau)  \right]_+ \;,
\end{equation}
where parameter $\alpha=1/30$ and $[.]_+$ vanishes for negative arguments. The microsaccade rate $r(t)$ was computed by averaging over microsaccades from all trials of a participant in a specific experimental condition. The resulting MSR as a function of time is qualitatively similar the time-courses of pupil size or event-related potentials, which are all stimulus-locked, continuous response functions averaged over many experimental trials. 

We identified six relevant time windows for our analyses:
Baseline ($-300$ to 0~ms), Target Onset (0 to 300~ms), Target Offset (300 to 600~ms), Return (600 to 900~ms), and Stability, which we divided into two 300~ms windows for consistency (900 to 1200~ms and 1200 to 1500~ms). The 300~ms window size is consistent with the finding that the MSR decreases in response to a target onset has a duration of approximately 300~ms before returning to the baseline \citep{Engbert2003}.

For the statistical analysis, we apply the knowledge that microsaccades are Poisson-distributed \citep{Engbert2006} and  conduct Poisson rate tests in each time window. In this analysis we compare the conditions Self--Other, and Strangers--Peers and report the ratio of the estimated Poisson rate. If the estimated Poisson rate is the the same in both groups the ratio will be 1, indicating no difference in MSR. 

\subsection{Relating Microsaccades and Pupil Size}
As both MSR and pupil size are reactive to our face conditions, in an explorative analysis we pose the question: are the two measures related, i.e.,  do trials with many microsaccades also show reduced dilation and vice versa? A relationship between the two would be an indication of a shared origin. The microsaccade effect precedes the pupil effect in time, with the drop in  MSR occurring between 50 and 100~ms after target onset while pupil size effects peak only after 1000~ms after target onset. In order to analyse the relationship of microsaccades and pupil size, we  begin by selecting the most diagnostic window for identity using microsaccades (i.e.,  the Target Offset Phase, 300-600~ms after target onset) and use it as a predictor for pupil size.  The distribution of microsaccade counts in this subset shows that the majority of trials (7919~trials) have a count of 0 microsaccades. 1992 trials had 1 or more saccades. This is a significant difference in sample size and may influence the results. We use a binary variable which encodes the presence of one or more microsaccades in the diagnostic window as an ad-hoc predictor variable of pupil size in an explorative LMM using the formula

\begin{equation}
Pupil\_Size \sim 1 + MS\_in\_phase*Face + (1|VP).
\end{equation}

The selection of the random effects structure is detailed in the appendix.

\section{Results}
In the present study we investigate the effect of seeing one's own face compared to familiar (Peers) and unfamiliar (Strangers) faces on involuntary eye measures. A subset of faces, including the Self condition, was shown repeatedly in order to explore how stimulus repetition influences the effect and to control for novelty effects. Repetition of unfamiliar faces also exposes a fourth recognition condition: recognition of a stranger's face.

\subsection{Pupil size}
\begin{figure}
    \centering
    \unitlength1mm
    \begin{picture}(150,90)
    \put(15,0){\includegraphics[width=0.8\textwidth]{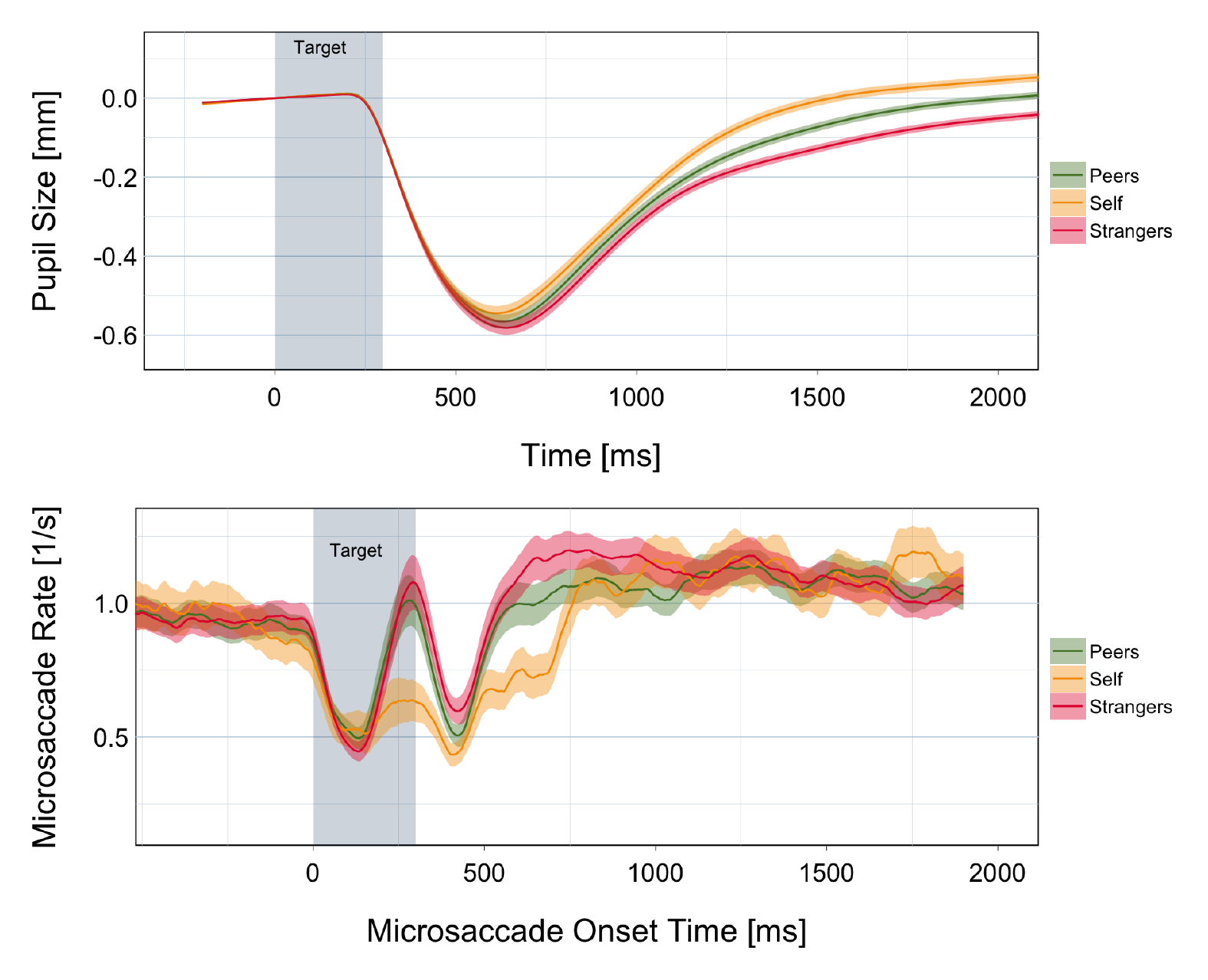}}
    \put(0,10){B}
    \put(0,65){A}
    \end{picture}
    
    \caption{\textbf{Pupil Size and Microsaccade rate over time.} (A) The lines show the change in pupil size normalized to the size at stimulus onset. The thin ribbons represent the between-subject standard error. (B) The lines show the rate of microsaccades after face onset (time zero), calculated for each participant and then averaged across participants for each condition. The shading show the between-subject standard error. The area highlighted in gray shows the time in which the target face was present on the screen.}
    \label{fig:ms_cat}
\end{figure}

In Figure \ref{fig:ms_cat}B as well as Figure \ref{fig:pupil_cat}, the average pupil reaction for each of the 3 face conditions reveals that there is a qualitative difference between all three. In the following we report the results of the LMM in each phase (refer also to Table \ref{tab:pupillmm}; the coefficients in the Table may be interpreted as the absolute difference in response in mm).

In the Baseline condition, as expected, the face conditions behave identically (Figure \ref{fig:pupil_cat} B). This is confirmed by the LMM, as presented in Table \ref{tab:pupillmm}.  The only significant effect is a minimal dilation trend over time.
The following window represents the initial constriction, which is a reflexive response to the stimulus presentation. In the constriction phase, too, no significant differences for the main effects can be found, with the exception of the comparison between the 4th and 5th presentation. 
In all following time windows, i.e.,  Dilation, Late Dilation and Stability Phases, we observe significant differences between the conditions. The effect size is largest in the Late Dilation phase for the Self--Others comparison and in the Stability phase for the Peers--Strangers comparison.
We conclude that starting at 600 ms after target onset recognition, and particularly self-recognition modulates pupil size. 

The repetition main effect is coded by a sliding differences contrast, meaning we compare subsequent presentations. We find a large number of the comparisons to be significant, showing that the pupil response becomes attenuated over repetitions. Note that we investigate the difference of each presentation to the next; a different contrast (e.g., a treatment contrast, comparing each level to the first repetition) would have likely resulted in more consistently significant effects. It may also be interesting to point out that while the effect becomes more muted over the presentations, the last comparison (5th vs. 4th presentation) actually indicates a reversal, perhaps pointing to an attenuation limit.

The interaction terms show a varying pattern of significance. The interactions with the Self--Others comparisons is consistently negative or absent, with the exception of the 2nd-3rd comparison. The interaction term can be interpreted as the attenuation of the pupil-dilation effect over repetitions. A negative term indicates that the attenuation in response to the Self condition is stronger than in the effect in the Peers and Strangers conditions.
The same interpretation can be applied to the Peers--Strangers and Repetition interactions. Here, a negative term means stronger attenuation in the Peers condition. Note that in Figure \ref{fig:pupil_cat} (e.g.,  in the Stability phase) it is evident that the second presentation of stranger produces a stronger dilation than the first (see 2nd--1st : Peers--Strangers interaction), presenting a contrast to the repetition muting effect found in the Peers and Self conditions. This is most likely caused by recognition of the face, when it it shown for the second time, i.e.,  unfamiliar face recognition.
In order to understand the full extent of these effects please refer to Figure \ref{fig:pupil_cat}.

\begin{table}[p]
\centering
\caption{\textbf{Results of pupil size LMMs.} The analysis was conducted using the same model for 5 time windows: Baseline (-50 -- 50~ms), Constriction (200 -- 600~ms), Dilation (600 -- 1200~ms), Late Dilation (600 -- 1200~ms), and Stability (2000 -- 3000~ms). The t values that are marked in bold and red are considered significant.\label{tab:pupillmm}}
\scalebox{0.7}{
\begin{tabular}{l|rrr|rrr}
\toprule
 & \multicolumn{3}{c}{Baseline} & \multicolumn{3}{c}{Constriction} \\
 & Estimate & Std. Error & t value & Estimate & Std. Error & t value \\
Contrast &  &  &  &  &  &  \\
\midrule
Intercept & -0.000057 & 0.000081 & -0.6984 & -0.295247 & 0.011274 & \color[HTML]{CB3D48} \bfseries -26.1884 \\
Self - Others (Se-O) & -0.000067 & 0.000161 & -0.4160 & 0.003681 & 0.004177 & 0.8814 \\
Peers - Strangers (P:St) & 0.000056 & 0.000100 & 0.5603 & 0.003096 & 0.001625 & 1.9052 \\
1st - 2nd Repetition & -0.000251 & 0.000215 & -1.1668 & -0.005188 & 0.004919 & -1.0548 \\
2nd - 3rd Repetition & 0.000097 & 0.000237 & 0.4079 & 0.001360 & 0.005532 & 0.2459 \\
3rd - 4th Repetition & 0.000198 & 0.000236 & 0.8377 & -0.000015 & 0.001259 & -0.0123 \\
4th - 5th Repetition & 0.000064 & 0.000207 & 0.3106 & 0.003244 & 0.001103 & \color[HTML]{CB3D48} \bfseries 2.9403 \\
\midrule
(Se-O) : (1st-2nd) & -0.000232 & 0.000578 & -0.4023 & -0.018211 & 0.003167 & \color[HTML]{CB3D48} \bfseries -5.7507 \\
(P:St) : (1st-2nd) & -0.000068 & 0.000167 & -0.4092 & -0.007136 & 0.000916 & \color[HTML]{CB3D48} \bfseries -7.7945 \\
(Se-O) : (2nd-3rd) & 0.000148 & 0.000590 & 0.2512 & 0.020158 & 0.003188 & \color[HTML]{CB3D48} \bfseries 6.3232 \\
(P:St) : (2nd-3rd) & 0.000217 & 0.000227 & 0.9581 & 0.010567 & 0.001219 & \color[HTML]{CB3D48} \bfseries 8.6705 \\
(Se-O) : (3rd-4th) & 0.000096 & 0.000587 & 0.1643 & -0.001853 & 0.003127 & -0.5928 \\
(P:St) : (3rd-4th) & -0.000204 & 0.000227 & -0.9004 & 0.000771 & 0.001207 & 0.6391 \\
(Se-O) : (4th-5th) & 0.000128 & 0.000473 & 0.2701 & -0.011617 & 0.002514 & \color[HTML]{CB3D48} \bfseries -4.6202 \\
(P:St) : (4th-5th) & 0.000169 & 0.000227 & 0.7460 & -0.002896 & 0.001206 & \color[HTML]{CB3D48} \bfseries -2.4026 \\
\midrule
Time & 0.000057 & 0.000001 & \color[HTML]{CB3D48} \bfseries 45.8196 & -0.001755 & 0.000002 & \color[HTML]{CB3D48} \bfseries -971.5124 \\
Trial & -0.000001 & 0.000002 & -0.5786 & 0.000062 & 0.000008 & \color[HTML]{CB3D48} \bfseries 7.3182 \\
\bottomrule
\end{tabular}

}

\scalebox{0.7}{
\begin{tabular}{l|rrr|rrr|rrr}
\toprule
 & \multicolumn{3}{c}{Dilation} & \multicolumn{3}{c}{Late Dilation} & \multicolumn{3}{c}{Stability} \\
 & Estimate & Std. Error & t value & Estimate & Std. Error & t value & Estimate & Std. Error & t value \\
Contrast &  &  &  &  &  &  &  &  &  \\
\midrule
Intercept & -0.383362 & 0.015263 & \color[HTML]{CB3D48} \bfseries -25.1169 & -0.067456 & 0.009939 & \color[HTML]{CB3D48} \bfseries -6.7872 & 0.032116 & 0.009002 & \color[HTML]{CB3D48} \bfseries 3.5677 \\
Self - Others (Se-O) & 0.052051 & 0.008528 & \color[HTML]{CB3D48} \bfseries 6.1036 & 0.095246 & 0.011187 & \color[HTML]{CB3D48} \bfseries 8.5140 & 0.072420 & 0.011285 & \color[HTML]{CB3D48} \bfseries 6.4171 \\
Peers - Strangers (P:St) & 0.008944 & 0.002811 & \color[HTML]{CB3D48} \bfseries 3.1818 & 0.012648 & 0.003699 & \color[HTML]{CB3D48} \bfseries 3.4190 & 0.013186 & 0.004262 & \color[HTML]{CB3D48} \bfseries 3.0939 \\
1st - 2nd Repetition & -0.001267 & 0.008189 & -0.1548 & -0.028325 & 0.009303 & \color[HTML]{CB3D48} \bfseries -3.0449 & -0.044097 & 0.011878 & \color[HTML]{CB3D48} \bfseries -3.7124 \\
2nd - 3rd Repetition & -0.006528 & 0.009706 & -0.6725 & -0.023712 & 0.010525 & \color[HTML]{CB3D48} \bfseries -2.2529 & 0.002909 & 0.012222 & 0.2380 \\
3rd - 4th Repetition & -0.008203 & 0.001374 & \color[HTML]{CB3D48} \bfseries -5.9713 & -0.014713 & 0.001434 & \color[HTML]{CB3D48} \bfseries -10.2633 & -0.022751 & 0.001483 & \color[HTML]{CB3D48} \bfseries -15.3451 \\
4th - 5th Repetition & -0.001765 & 0.001204 & -1.4660 & 0.011255 & 0.001257 & \color[HTML]{CB3D48} \bfseries 8.9562 & 0.011447 & 0.001300 & \color[HTML]{CB3D48} \bfseries 8.8071 \\
\midrule
(Se-O) : (1st-2nd) & -0.031241 & 0.003458 & \color[HTML]{CB3D48} \bfseries -9.0339 & -0.092357 & 0.003609 & \color[HTML]{CB3D48} \bfseries -25.5899 & -0.140653 & 0.003733 & \color[HTML]{CB3D48} \bfseries -37.6795 \\
(P:St) : (1st-2nd) & -0.010495 & 0.001000 & \color[HTML]{CB3D48} \bfseries -10.4924 & -0.027946 & 0.001044 & \color[HTML]{CB3D48} \bfseries -26.7681 & -0.025951 & 0.001080 & \color[HTML]{CB3D48} \bfseries -24.0306 \\
(Se-O) : (2nd-3rd) & 0.023121 & 0.003481 & \color[HTML]{CB3D48} \bfseries 6.6426 & 0.010414 & 0.003633 & \color[HTML]{CB3D48} \bfseries 2.8670 & 0.038542 & 0.003757 & \color[HTML]{CB3D48} \bfseries 10.2586 \\
(P:St) : (2nd-3rd) & 0.002437 & 0.001331 & 1.8316 & 0.002735 & 0.001389 & 1.9698 & 0.007104 & 0.001436 & \color[HTML]{CB3D48} \bfseries 4.9462 \\
(Se-O) : (3rd-4th) & 0.001121 & 0.003412 & 0.3287 & 0.000110 & 0.003561 & 0.0310 & -0.012940 & 0.003683 & \color[HTML]{CB3D48} \bfseries -3.5137 \\
(P:St) : (3rd-4th) & 0.004935 & 0.001317 & \color[HTML]{CB3D48} \bfseries 3.7478 & 0.008294 & 0.001374 & \color[HTML]{CB3D48} \bfseries 6.0358 & 0.011110 & 0.001421 & \color[HTML]{CB3D48} \bfseries 7.8176 \\
(Se-O) : (4th-5th) & -0.030656 & 0.002744 & \color[HTML]{CB3D48} \bfseries -11.1721 & -0.038631 & 0.002864 & \color[HTML]{CB3D48} \bfseries -13.4906 & -0.039241 & 0.002962 & \color[HTML]{CB3D48} \bfseries -13.2499 \\
(P:St) : (4th-5th) & -0.003351 & 0.001316 & \color[HTML]{CB3D48} \bfseries -2.5474 & -0.003356 & 0.001373 & \color[HTML]{CB3D48} \bfseries -2.4441 & -0.009235 & 0.001420 & \color[HTML]{CB3D48} \bfseries -6.5038 \\
\midrule
Time & 0.000754 & 0.000001 & \color[HTML]{CB3D48} \bfseries 573.5500 & 0.000201 & 0.000001 & \color[HTML]{CB3D48} \bfseries 195.0739 & 0.000068 & 0.000001 & \color[HTML]{CB3D48} \bfseries 80.0736 \\
Trial & 0.000028 & 0.000009 & \color[HTML]{CB3D48} \bfseries 3.0541 & -0.000633 & 0.000010 & \color[HTML]{CB3D48} \bfseries -65.6442 & -0.000889 & 0.000010 & \color[HTML]{CB3D48} \bfseries -89.1710 \\
\bottomrule
\end{tabular}

}

\end{table}





\begin{figure}
    \centering
    \begin{picture}(300, 500)(0, 0)
    \put(0, 0){\includegraphics[width=0.81\textwidth]{Figure3.pdf}}
    \put(2, 380){A}
    \put(2, 260){B}
    \put(200, 260){C}
    \put(2, 135){D}
    \put(200, 135){E}
    \put(2, 12){F}
\end{picture}

    \caption{\textbf{Pupil size over time.} The top panel (A) shows the average measured pupil size for each category of images. The colored background separates the response signature into 5 time windows: Baseline (B, red), Constriction (C, yellow), Dilation (D, rose), Late Dilation (E, green), and Stability (F, blue). Panel A shows that the main effect of pupil size is evident from the Dilation phase onward, where the pupil size becomes significantly different for Strangers, Peers, and Self. \textcolor{black}{The bottom panels, B, C, D, E, and F, shows a detailed view of the pupil size over time in the respective time window. Each line in a range of shades of gray shows one repetition, with black being the first viewing of the face and white the 10\textsuperscript{th}.}}
    \label{fig:pupil_cat}
\end{figure}

\begin{figure}
    \centering
    \includegraphics[width=0.8\textwidth]{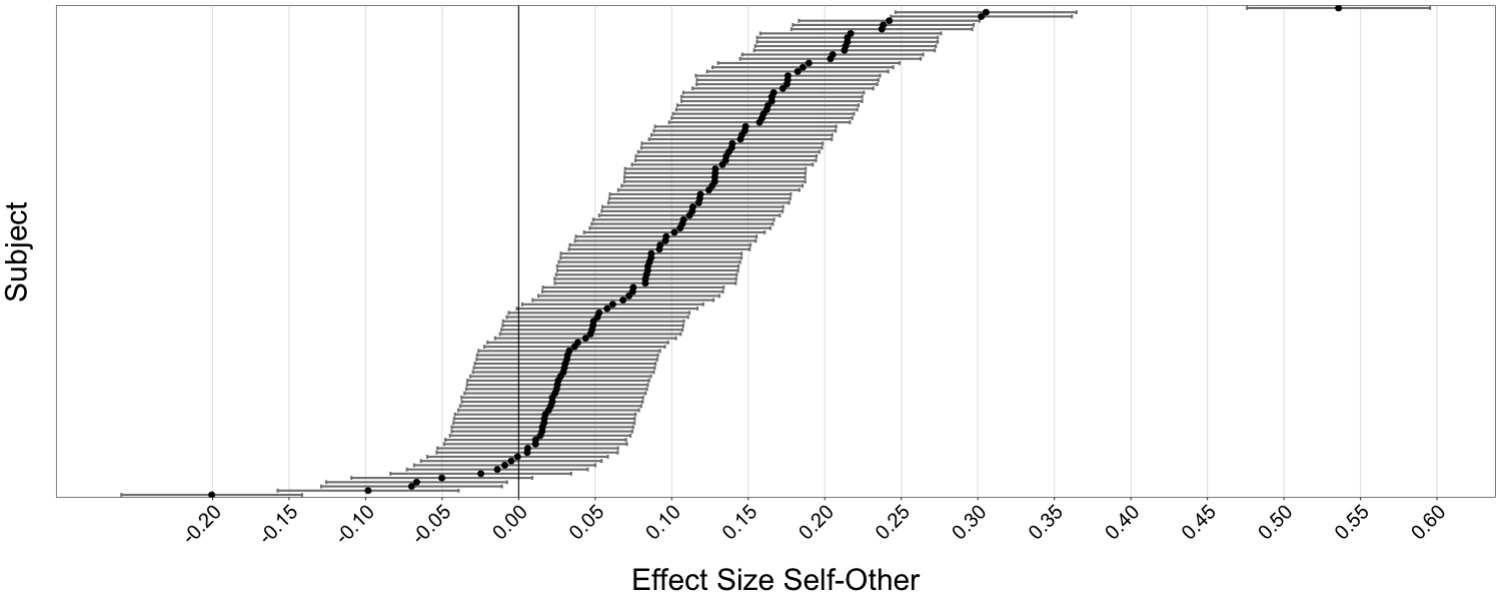}
    \caption{\textbf{Effect size of the Self-Others comparison, by participant.} Mean values are computed from the estimated Random Effect Terms. Zero in this plot means that there was no effect, negative values indicate a reversal, i.e., reacting less strongly to Self than to Others. The great majority of participants show the effect (to the right of the zero line). A small subset shows a reversed effect.} 
    \label{fig:subj_effect}
\end{figure}

As a further analysis, we were interested in how consistently participants were exhibiting the effect. Using the terms computed by the LMM, we compare the effect size of the Self-Others comparison for individual subjects, as estimated by the random effects in the LMM. The results are plotted in Figure \ref{fig:subj_effect}. We find that the majority of subjects shows the effect in the expected direction. Some individuals show only a very minor expression of the effect and some even mildly reverse the trend, i.e., exhibit less dilation in response to their own face instead of more.

\subsubsection{Microsaccades}


We calculated the MSR for each subject and condition over time (Figure \ref{fig:ms_cat}).
The time course of microsaccades is best described as two interacting processes. First, simple display changes cause an inhibition of microsaccades \citep{Engbert2003}. As our paradigm involved two changes in display (image onset and image offset) both display changes independently cause the MSR to drop. 
\textcolor{black}{Second, microsaccades can also be inhibited by visual on- or offsets and can be modulated by  cognitive factors such as exogeneous or endogeneous shifts of attention \citep{Engbert2006a}. In a computational model, \citet{Engbert2012} suggested that the combination of two processes, (i) the modulation of a threshold for triggering of microsaccades and (ii) a transient reduction of microsaccades, can explain most of the experimental findings for lower-level display changes and higher-level cognitive task manipulations. An alternative model suggests that microsaccades are modulated via phase resetting of neural oscillations during a stimulus onset \citep{Hafed2013}, which is further discussed in the context of microsaccadic inhibition \citep{Buonocore2023,Hafed2021}.}


As suggested by previous research, we find that the MSR responds strongly to display changes. An inhibition of microsaccades is clearly  visible in the rapid drop in MSR at image onset and image offset. Typically the MSR recovers quickly, rising to the baseline level. Figure \ref{fig:ms_cat} shows that in the Self condition microsaccades are inhibited for a longer period of time. In all other conditions, the rate recovers to the baseline level before the second inhibition occurs at target offset. In the Self condition, the second inhibition occurs while the first is still active, leading to a further drop in the MSR and a recovery that is correspondingly longer than in the other conditions. Thus the average time-course of the MSR is qualitatively different in the Self condition. Thus, our results are compatible with earlier findings on microsaccadic inhibition in the oddball paradigm \citep{Valsecchi2006}.

Table \ref{tab:ms_tests} shows the results of the Poisson rate tests.  The Estimate should be interpreted as the ratio of the Poisson rates in the compared conditions, e.g.,  the estimate of $1.23$ in the Return Phase when comparing Self and Others means that the rate in the Others condition is higher by a factor of 1.23 than in the Self Condition. We find significant differences between the Self and Others conditions, as well as between Peers and Strangers in the period between 300~ms and 900~ms.

\begin{table}[p]
\centering
\caption{\textbf{Results of the Microsaccade Poisson rate tests.} The test compared the Poisson rate of two groups. P-values marked in red are considered significant. The t values marked in bold and red are considered significant.}
\label{tab:ms_tests}
\begin{tabular}{lllrrrr}
\toprule
Contrast & Window & Time & Estimate & CI1 & CI2 & p-value \\
\midrule
Self - Others & Baseline & -300--0 ms & 1.0187 & 0.9072 & 1.1471 & 0.7922 \\
Self - Others & Target Onset & 0--300 ms & 1.1265 & 0.9759 & 1.3063 & 0.1094 \\
Self - Others & Target Offset & 300--600 ms & 1.3925 & 1.2066 & 1.6145 & \color[HTML]{CB3D48} \bfseries 0.0000 \\
Self - Others & Return & 600--900 ms & 1.2274 & 1.0942 & 1.3808 & \color[HTML]{CB3D48} \bfseries 0.0003 \\
Self - Others & Stability 1 & 900--1200 ms & 0.9889 & 0.8892 & 1.1024 & 0.8279 \\
Self - Others & Stability 2 & 1200--1500 ms & 0.9841 & 0.8860 & 1.0955 & 0.7674 \\
\midrule
Peers - Strangers & Baseline & -300--0 ms & 1.0235 & 0.9513 & 1.1012 & 0.5299 \\
Peers - Strangers & Target Onset & 0--300 ms & 0.9457 & 0.8671 & 1.0313 & 0.2047 \\
Peers - Strangers & Target Offset & 300--600 ms & 1.1052 & 1.0218 & 1.1956 & \color[HTML]{CB3D48} \bfseries 0.0118 \\
Peers - Strangers & Return & 600--900 ms & 1.1111 & 1.0395 & 1.1877 & \color[HTML]{CB3D48} \bfseries 0.0018 \\
Peers - Strangers & Stability 1 & 900--1200 ms & 1.0445 & 0.9759 & 1.1180 & 0.2097 \\
Peers - Strangers & Stability 2 & 1200--1500 ms & 1.0327 & 0.9656 & 1.1046 & 0.3496 \\
\bottomrule
\end{tabular}

\end{table}

\subsection{Relationship of Microsaccades and Pupil Size}
We calculated a model using the presence of microsaccades in the most diagnostic window, as found by the previous analysis, as a predictor for pupil size.  The result of the LMM is presented in Table \ref{tab:pupilmscorr}. First, consistent with the LMMs of pupil size alone, we find a significant effect of both Self--Others and Peers--Strangers in the Dilation, Late Dilation and Stability time widows. Second, a predictive effect of microsaccades is present only in the earlier phases of the pupil response (constriction, dilation). 

\begin{table}[p]
\centering
\caption{\textbf{Estimated terms of the LMM relating microsaccades and pupil size.} Occurrence of microsaccades in the most diagnostic window (300–600~ms) is used as a predictor for pupil size. The predictive effect of microsaccades can only be found in the early phases of the pupil response. The t values that are marked in bold and red are considered significant.}
\label{tab:pupilmscorr}
\scalebox{0.7}{
\begin{tabular}{l|rrr|rrr|}
\toprule
Contrast & \multicolumn{3}{c}{Baseline} & \multicolumn{3}{c}{Constriction} \\
 & Estimate & Std. Error & t value & Estimate & Std. Error & t value \\
\midrule
Intercept & -0.000364 & 0.000097 & \color[HTML]{CB3D48} \bfseries -3.7492 & -0.294699 & 0.011040 & \color[HTML]{CB3D48} \bfseries -26.6932 \\
Self - Others (Se-O) & -0.000143 & 0.000194 & -0.7366 & 0.007281 & 0.003744 & 1.9444 \\
Peers - Strangers (P:St) & -0.000022 & 0.000065 & -0.3385 & 0.002272 & 0.001248 & 1.8195 \\
MS - No MS & 0.000227 & 0.000186 & 1.2207 & -0.008003 & 0.003669 & \color[HTML]{CB3D48} \bfseries -2.1812 \\
(Se-O) : (MS - No MS  & 0.000456 & 0.000492 & 0.9266 & -0.007382 & 0.009573 & -0.7711 \\
(Se-O) : (MS - No MS  & 0.000136 & 0.000141 & 0.9656 & 0.001129 & 0.002758 & 0.4092 \\
\bottomrule
\end{tabular}

}
\scalebox{0.7}{
\begin{tabular}{l|rrr|rrr|rrr|}
\toprule
Contrast & \multicolumn{3}{c}{Dilation} & \multicolumn{3}{c}{Late Dilation} & \multicolumn{3}{c}{Stability} \\
 & Estimate & Std. Error & t value & Estimate & Std. Error & t value & Estimate & Std. Error & t value \\
\midrule
Intercept & -0.378145 & 0.014605 & \color[HTML]{CB3D48} \bfseries -25.8907 & -0.058821 & 0.009930 & \color[HTML]{CB3D48} \bfseries -5.9236 & 0.034671 & 0.008973 & \color[HTML]{CB3D48} \bfseries 3.8638 \\
Self - Others (Se-O) & 0.049080 & 0.006401 & \color[HTML]{CB3D48} \bfseries 7.6677 & 0.083602 & 0.007790 & \color[HTML]{CB3D48} \bfseries 10.7321 & 0.066590 & 0.008908 & \color[HTML]{CB3D48} \bfseries 7.4750 \\
Peers - Strangers (P:St) & 0.013208 & 0.002134 & \color[HTML]{CB3D48} \bfseries 6.1889 & 0.027017 & 0.002597 & \color[HTML]{CB3D48} \bfseries 10.4025 & 0.025628 & 0.002970 & \color[HTML]{CB3D48} \bfseries 8.6292 \\
MS - No MS & -0.014450 & 0.006271 & \color[HTML]{CB3D48} \bfseries -2.3042 & -0.010262 & 0.007622 & -1.3464 & -0.004084 & 0.008705 & -0.4691 \\
(Se-O) : (MS - No MS  & -0.009118 & 0.016365 & -0.5571 & -0.002875 & 0.019914 & -0.1444 & 0.019757 & 0.022770 & 0.8677 \\
(Se-O) : (MS - No MS  & -0.004482 & 0.004715 & -0.9507 & -0.006422 & 0.005737 & -1.1194 & -0.006034 & 0.006559 & -0.9199 \\
\bottomrule
\end{tabular}

}

\end{table}
\color{black}
\section{Discussion}
In a large-scale eye-tracking experiment, we investigated the behavioral correlates of face recognition in involuntary eye movements. We recorded pupil size and fixational eye movements in two groups of participants. Each participant saw their own face (Self), the familiar faces of members of their own peer group (Peers), or unfamiliar faces (Strangers). Selected faces, including the Self condition, were repeated to investigate familiarization effects. This study design provides a well-controlled data set, where, across participants, each individual face appeared in each condition (Peers, Strangers, Self). We observe that face recognition, particularly self-recognition, modulates the response signatures of both pupil dilation and MSR.


\subsection{Recognition of Familiar Faces}
First, we compared the recognition of familiar faces to completely unfamiliar faces, i.e. Peers vs. Strangers. Recognizing a familiar faces is related to a stronger and longer-lasting inhibition of microsaccades, starting around 200~ms after target onset. Recognition effects on MSR were first documented by \citet{Rosenzweig2019,Rosenzweig2020}, however a study by \citet{Chen2023} did not reproduce this result. We also observed increased pupil dilation in response to familiar faces starting around 600~ms after target onset. Our findings regarding pupil dilation are in agreement with previous work \citep{Chen2023}. 
\textcolor{black}{In accordance with the literature, we propose that early effects (microsaccade response, early stages of pupil response), are related purely to recognition and saliency \citep{Mathot2018,Larsen2018}. Early recognition responses in the IT cortex occur 50-100~ms after target onset \citep{Sugase1999, Perrett1982} and SC shows object detection responses even earlier \citep{Nguyen2012, Nguyen2014, Nguyen2016, Bogadhi2023}. The data show effects on microsaccade inhibition less than 100~ms later, suggesting that feedback from the fusiform face area (FFA) may have a direct influence on the oculomotor system.}
The mid and late stages of the pupil response are likely related more to arousal and emotional responses \citep{Mathot2018}.

\subsection{Self-Recognition}
Similar effects as in the Peers--Strangers comparison are evident in the Self--Others comparison, albeit significantly larger in scale.
We hypothesized that the faces of strangers capture little attention and are grouped together, while one's own face is qualitatively different, even to familiar faces.
We find a strong individual variability in the effect, which, we speculate, may be related to differences in the emotional response related to self-image.
It is important to note that the Self--Others comparison is not perfectly balanced regarding stimulus frequency. While many different Peers and Strangers were shown, there is only ever one Self face, which is repeated. We controlled for repetition by also repeating certain Peers' and Strangers' faces. However, the relative frequency of Self images was still lower than of the Others images. 

We find the largest effects for the first presentation of Self, and an attenuation over repetitions.
Thus, it is likely that genuine self-recognition effects are further enhanced by a stimulus-probability effect, similar to the Oddball paradigm.



\subsection{Repetition Effects and Recognition of Unfamiliar Faces}
Repetitions of the same image led to reduced pupil dilation. We observed a strong attenuation of the effect over repetitions, such that the 10th Self-presentation is comparable to the Strangers response (see Figure \ref{fig:pupil_cat}). The strongest repetition effects are visible in the Late Dilation and Stability Phases. Unlike the neurophysiological novelty response which occurs 300~ms after target onset (compare: P300 component), this novelty effect on pupil constriction begins much later, around 1,000~ms.

On its own, the attenuation over repetitions seems discouraging for real-world applications, where pictures may be presented repeatedly. However, this is not true of the early response: in the dilation phase, the faster and stronger dilation in the Self condition is still evidently different from the average Strangers condition. Whether this effect is short- or long-term, i.e.,  whether it would persist over multiple sessions remains to be investigated.

The first response to seeing one's own face is comparatively much larger than in any other condition, and is much reduced in subsequent repetitions   (see Figure \ref{fig:pupil_cat}, or "(Se-O) : (1st-2nd)" in Stability phase in Table \ref{tab:pupillmm}).
Moreover, when a stranger's face is repeated, i.e., during the first \textit{recognition} of a now-familiarized stranger, the dilation is stronger than during the very first viewing (see Figure \ref{fig:pupil_cat}, or "(F:St) : (1st-2nd)" in Late Dilation phase in Table \ref{tab:pupillmm}).
This outlines the effect of the pure recognition of a face, in the absence of any emotional reaction. Further recognitions of the stranger reduce the effect in accordance with the repetition effect.

\subsection{The Relationship of Pupil Size and Microsaccade Rate}
We find that microsaccadic inhibition and the pupil dilation are only weakly correlated. Trials that have particularly few microsaccades do not serve well as a predictor for strong dilation. This observation is surprising, as we expected both measures to share a common cause. Our results suggest that MSR and pupil dilation are controlled independently and via separate pathways, although they commonly react to cognitive and emotional factors. The lack of correlation could confer a positive effect on models using a conjunction of both features for its prediction.

\subsection{Potential applications}
The question of facial familiarity is crucial in security applications.
Concealed information testing (CIT) serves to identify individuals with knowledge of a crime or perpetrator without relying on verbal statements.
The robust effects presented here serve as a good basis for applications in continuous authentication procedures or identity validation. Initial explorations show the feasibility of using these measures as traits for biometric classification \citep{graupner2023}. Possible real-world scenarios include the detection of face presentation attacks (e.g., video deepfakes) in online video conferences.
As we found an attentuation of the effect over image repetition, a remaining important consideration is whether the observed differences can be discerned at the level of individual trials and sustained over repetitions. More application-oriented studies will be required to validate the effectiveness of involuntary eye movements for practical use cases.

\section*{Acknowledgements}
The authors acknowledge support from Bundesdruckerei GmbH, Berlin, Germany.

\section*{Ethics Statement}
No animal studies are presented in this manuscript. Ethical approval was not required for the studies involving humans because the local legislation and institutional requirements do not require specific approval for studies where no sensitive or identifiable information is published.
The studies were conducted in accordance with the local legislation and institutional requirements. The participants provided their written informed consent to participate in this study in accordance with the General Data Protection Regulation of European Union. Specifically, under Article 8 of the GDPR, the processing of personal data of people over the age of 16  does not require consent by the holder of parental responsibility. 
No potentially identifiable images or data are presented in this study.

\bibliography{main.bib}

\appendix
\renewcommand{\thefigure}{\Alph{section}\arabic{figure}}
\renewcommand{\thetable}{\Alph{section}\arabic{figure}}
\setcounter{figure}{0}
\setcounter{table}{0}

\section{Supplemental Online Materials}
\subsection{LMM Random Effects Structure}
\label{appx:sec:lmm}
We used linear mixed models (LMMs) to analyze the pupillometry data in 5 time windows. In all time windows we employed the same fixed and random effect structure. Random effects were estimated for both participants and images, but mainly with the aim of correcting for their influence. We initially formulated a hypothesis-driven random effect structure, including random intercepts for both participants and images, and random slopes for each the Faces comparisons. We included the Repetition comparisons as random slopes for each subject (but not each image, as not all images were necessarily seen in each repetition). We also excluded random interaction slopes. Model reduction was then performed iteratively until convergence was achieved without issues. Specifically, correlation terms were first removed, followed by the least varying random effect terms. In cases where two converging models with different random effect terms were obtained, we used Bayesian-Information-Criterion (BIC) to select the best model. To ensure that none of the models were degenerate, a principal component analysis was performed on the random effect terms \citep{Bates2015a}. The final model structure  is as follows,
\begin{equation}
+ (1+Face+Rep2+Rep3|VP) + \\
 (1+Face(Se-O)|Img).
\end{equation}

We performed the same procedure for the random effects structure of the explorative LMM that explores the relationship between microsaccade occurrence and pupil size. Here we included random intercepts by subject. As the predictor variable was ad-hoc, group sizes were uneven and a random effect of image was excluded. After the model reduction procedure we arrived at a model which defines only a random intercept for each subject. 

\begin{equation}
+ (1|VP).
\end{equation}

\subsection{Controlling for Knowing}
\begin{figure}
    \centering
    \begin{picture}(500, 200)(0,0)
    \put(10, 0){\includegraphics[width=0.9\textwidth]{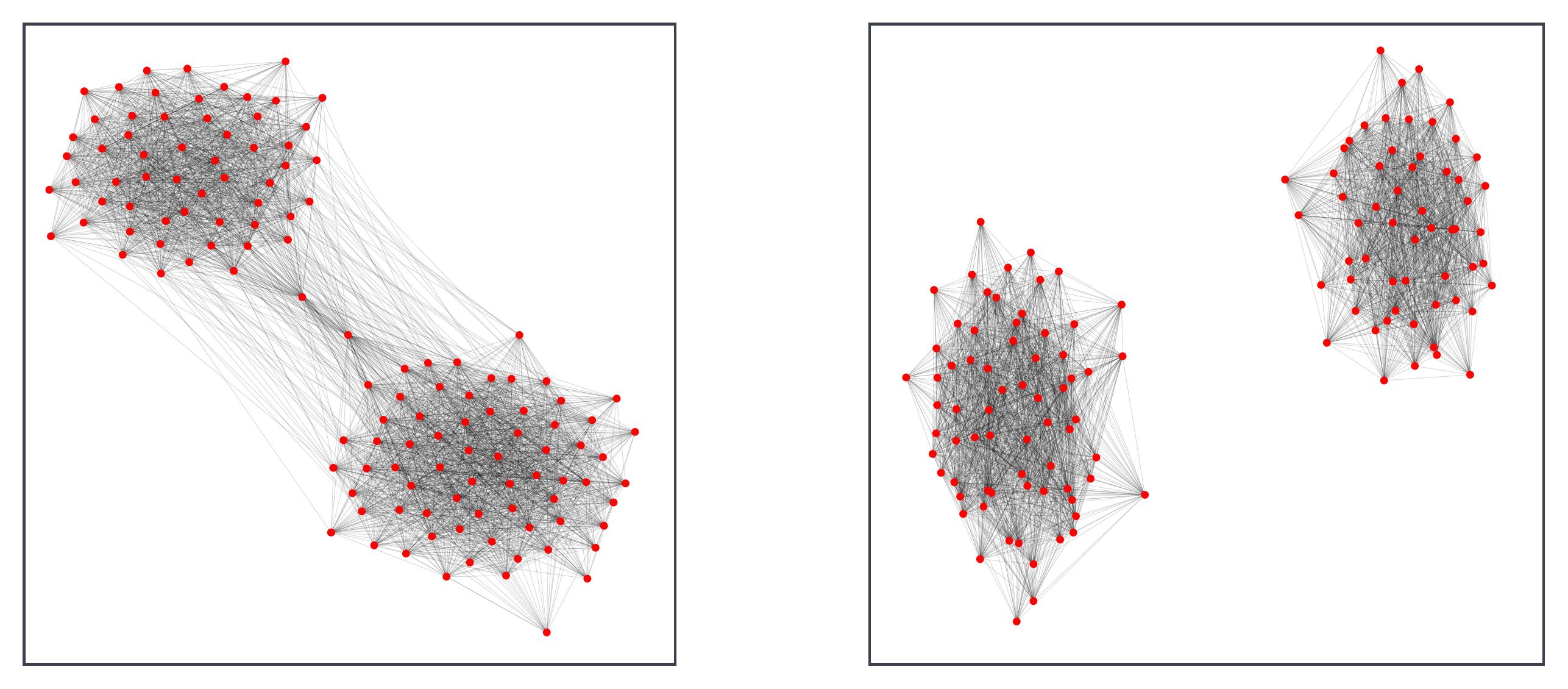}}
    \put(0, 7){A}
    \put(230, 7){B}
    \end{picture}

    \caption{\textbf{Social Network of participants.} The two clusters represent the two school classes. The right panel shows the social participants that knew each other. The left panel shows the connections after removing trials that violated the expectation that students only know people from their own class.}
    \label{fig:social_network}
\end{figure}

After the experiment, participants were shown all faces one more time. They were asked to indicate by pressing one of 3 buttons on a \emph{ViewPixx} button box whether they know the person in the picture, did not know them, or know them very well. With this information we can exclude trials with subjects who coincidentally did know each other across the schools. Figure \ref{fig:social_network} shows a visualization of the social network between the participants. This data also allows potential exploratory analyses using the distinction of close peers versus acquaintances.

\subsection{Controlling for Self Image}
For the purposes of an exploratory analysis, we asked three questions that were to be answered on a 5 point scale, i.e., 
\begin{enumerate}
\itemsep0ex
\item How did you like the photographs in general? \emph{original: Wie fanden Sie die Fotos allgemein?}
\item How did you like the photograph of yourself? \emph{original: Wie fanden Sie das Foto von sich selbst?}
\item How happy are you with your own appearance in general? \emph{original: Wie glücklich sind Sie mit Ihrem eigenen Erscheinungsbild allgemein?}
\end{enumerate}

\noindent Questions 1 and 2 were to be answered with
\begin{enumerate}
\itemsep0ex
\item Liked very much \emph{original: Sehr gelungen}
\item Liked \emph{original: Gelungen}
\item Neutral \emph{original: Durchschnittlich}
\item Did not like much \emph{original: Eher nicht gelungen}
\item Did not like \emph{original: Nicht gelungen}
\end{enumerate}

\noindent and question 3 with
\begin{enumerate}
\itemsep0ex
\item Very happy \emph{original: Sehr glücklich}
\item Happy \emph{original: Glücklich}
\item Neutral \emph{original: neutral}
\item Rather unhappy \emph{original: Eher unglücklich}
\item Unhappy \emph{original: unglücklich}
\end{enumerate}

\end{document}